\documentclass[preprintnumbers, prd, onecolumn, floatfix,
preprintnumbers,showpacs, letterpaper, superscriptaddress,nofootinbib]{revtex4}
\usepackage{graphicx,epsfig}
\usepackage{amsmath}
\usepackage{eurosym}
\usepackage{amssymb}
\usepackage{subfigure}
\usepackage{color}
\usepackage{amsfonts}
\usepackage[usenames,dvipsnames,svgnames]{xcolor}  
\def\be{\begin{equation}}
\def\ee{\end{equation}}
\def\barr{\begin{array}}
\def\earr{\end{array}}
\def\bea{\begin{eqnarray}}
\def\eea{\end{eqnarray}}
\def\bfig{\begin{figure}}
\def\efig{\end{figure}}

\usepackage[colorlinks]{hyperref}
\usepackage{hyperref}   
\hypersetup{colorlinks=true,linkcolor=beamer@PRD, citecolor=beamer@PRD}
\definecolor{beamer@PRD}{RGB}{46,48,146}

\begin{document}
\title{The quark chemical potential of QCD phase transition and the stochastic background of gravitational waves}

\author {\bf{Salvatore Capozziello} }

\email{capozziello@na.infn.it (Corresponding  Author)}

\affiliation{Dipartimento di Fisica "E. Pancini", Universit\`{a} di Napoli
  ``Federico II'', Complesso Universitario di Monte Sant'Angelo,
  Edificio G, Via Cinthia, I-80126, Napoli, Italy.}

  \affiliation{Istituto Nazionale di Fisica Nucleare (INFN) Sezione
  di Napoli, Complesso Universitario di Monte Sant'Angelo, Edificio G,
  Via Cinthia, I-80126, Napoli, Italy.}

 \affiliation{Gran Sasso Science Institute, Viale F. Crispi, 7, I-67100,
L'Aquila, Italy.}

\affiliation{Tomsk State Pedagogical University, ul. Kievskaya, 60, 634061 Tomsk, Russia.}

\author {\bf{Mohsen Khodadi} }

\email{mohsen.khodadi@gmail.com}

\affiliation{Department of Physics, Faculty of Basic Sciences,\\
University of Mazandaran, P. O. Box 47416-95447, Babolsar, Iran.}

\affiliation{Research Institute for Astronomy and Astrophysics of Maragha\\
(RIAAM), P. O. Box 55134-441, Maragha, Iran.}

\author {\bf{Gaetano Lambiase} }

\email{lambiase@sa.infn.it}

\affiliation{Dipartimento di Fisica E.R. Cainaiello, University of Salerno, Via
Giovanni Paolo
II, I 84084-Fisciano (SA), Italy.}

\affiliation{INFN, Gruppo Collegato di Salerno, Sezione di Napoli, Via Giovanni Paolo II, I
84084-Fisciano (SA), Italy.}

\date{\today}

\begin{abstract}

The detection of stochastic background of gravitational waves (GWs),  produced
by cosmological phase transitions (PTs),
is of fundamental importance because allows to probe the physics related to PT energy
scales. Motivated by the decisive role of non-zero quark chemical potential
towards understanding physics in the core of  neutron stars, quark stars
and heavy-ion collisions, in this paper we qualitatively explore the stochastic
background of GW spectrum generated by a cosmological source such as high-density QCD first order PT during the early Universe.
Specifically,
we calculate the frequency peak $f_{peak}$
redshifted at today time and the fractional energy density $\Omega_{gw}h^2$
in light of equation-of-state improved by the finite quark
(baryon) chemical potential (we consider an effective three flavor chiral quarks
model of QCD). Our calculations reveal a striking increase in $f_{peak}$ and $\Omega_{gw}h^2$
due to the quark chemical potential, which means to improve the chances of detection, in possible future observations (in particular SKA/PTA experiments), of
the stochastic background of GWs from QCD first order PT. Even if the improvements could be weak, by updating the sensitivity of  relevant detectors in the future, we can still remain hopeful.
Concerning the phenomenological contribution of QCD equation-of-state, and in particular the possibility to
detect a stochastic GW signal, we further show that the role of the quark chemical potential is model-dependent. This feature allows to
discriminate among  possible QCD effective models depending on their capability
to shed light on the dynamic of QCD-PT through future observations of primordial GWs.
In this perspective, the results are indeed encouraging to employ the GWs to study the QCD PT
in high density strong interaction matter.
\end{abstract}

\maketitle

\section{Introduction}

Electromagnetic radiation (EMR)-based observations, in particular the cosmic microwave
background (CMB) radiation emitted 380,000 years after Big Bang, are one of the most important
source of our current understanding of the Universe. This window on the early
Universe, however, has a limited horizon because, before that time, the Universe was opaque to EMR.
In  other words, any signals from earlier times can only be observed indirectly via
their footprints on the CMB which is a serious restriction in the sense that many important
questions in cosmology require information about the events during the first instant (nanoseconds and
microseconds) after the Big Bang. Even though there is a wide scope for exploiting the
entire electromagnetic spectrum, the above restriction remains in the face of very early Universe.
In this respect gravitational waves (GWs) can remove such a limitation and, therefore, could be able to provide a new
source of information about the primordial Universe. Because the GWs, like  electromagnetic waves,
travel at the speed of light owing to their extremely weak interaction with matter, the Universe
has always been transparent to them, providing hence a direct view of the  very first epochs
of the Universe. As a result, the combination of GW observations  \cite{ref:ligo1, ref:ligo2}
with  CMB observations allows to address some puzzle of present cosmology, such as the nature of dark
energy and dark matter (see for example \cite{addazi}), early evolution of structures, and so on\footnote{It is interesting to point out that, the physical meaning
of GWs, as the vibrations of spacetime predicted by Einstein,  was discussed at the Chapel Hill conference \cite{ref:1957}. In this regards,
it was understood that  GWs are energy carriers   passing through the spacetime  that affects the position of
particles in its path \cite{ref:Bondi1957, ref:Bondi1959}.}. Moreover, the existence of
GWs is not only restricted to General Relativity, but indeed they can be found in many modified theories of
gravity \cite{Report,Report2,astrop, sergei,Calmet,lambiase1,lambiase2}.

Generally, GWs emerge from cosmological and relativistic astrophysical sources. Concerning the astrophysical sources,
coalescing binary neutron stars, as demonstrated, are regarded as the most likely GW sources 
to be seen by VIRGO/LIGO interferometers. Some researches suggest that  GWs emitted by the merging
of  binary neutron stars could yield significant information about the equation-of-state (EoS) of dense matter and the related gravitational theory, see e.g., \cite{Faber,Taniguchi,Dorota,Oechslin,Bejger,felix}. As recently discussed in \cite{Soroush},
GWs  produced  by
binary mergers lead to circular polarization signal generation which may be captured by X-ray polarimetry missions.
The detection of GWs with cosmological origins (unlike their astrophysical counterpart have a stochastic and random character)
has special significance in the sense that it can be
physically relevant for the Universe evolution in the very early stages. Concerning the cosmological origin,
there are primordial GWs emerging from some processes as inflation and
reheating epochs. Remarkably, these GWs can be tracked and measured via their unique footprint on the CMB.
However, during the expansion of the Universe after inflation, and according to the Standard Model of elementary physics,
we have to take into account also for phase transitions (PTs) at lower temperatures that can generate GWs:
the electro-weak and quantum chromodynamics (QCD) PTs. Despite the fact that there is still no final agreement on the type of these two PTs, it is believed
that GWs should be produced in models with enough long duration,  i.e. during first order PTs. More technically, in the first
order cosmological PT-based GWs, the nucleation of bubbles is due to a series of expansions and encounters with each
other, resulting in a major stochastic background of GWs. The eLISA interferometer \cite{ref:elisa}, as a space-based GW detector
as well as other operators as PTA \cite{ref:pta2013} and SKA \cite{ref:ska2004} with different
sensitivities\footnote{It should be noted that GWs cover a wide range of frequencies that, for identification of each, require particular technology.
For instance, relevant frequency of QCD-PT-based stochastic GWs is around $<10^{-5}$HZ. In Refs. \cite{ref:Kumar2017,noi},
frequencies related to other GW sources are discussed.}, are designed to trap stochastic background of GWs arising from electro-weak and QCD PTs,
respectively. Note that, similarly to the CMB radiation which is emitted from the last scattering surface, the stochastic background of GWs is produced from
distant surfaces of the Universe perimeter, at the PTs epoch \cite{ref:Geller2018}.

The PT, that we  are considering in this paper as a cosmological source of stochastic GWs,
is related to the QCD epoch\footnote{Recall that, in high energy regime, QCD is an asymptotic freedom and perturbative theory while, at low energy, it is strongly-coupled so that a perturbative approach is not useful \cite{ref:Ahmad2018}.}.  After a few microseconds from the Big Bang, a PT happened  from a mixed phase of quark-gluon plasma (QGP) evolving in hadrons. To achieve a real understanding of QCD PT dynamics,
without an appropriate thermodynamics related to a relevant equation-of-state (EoS), is impossible. In this respect to have
a proper EoS becomes therefore crucial, especially when the stochastic background of GWs generated by any PT is known. It severely
depends on the  critical temperature\footnote{In \cite{noi}, it is shown, in
the Planck physics extended framework, assuming  some natural cutoffs on the length and momentum
of particles into QCD thermodynamic,  that the stochastic GW background results  affected even in the absence of
change in the critical temperature.} $T_*$ (in particular here $T_{*(QCD)}\sim$ a few hundred MeV), \cite{ref:Kumar2017,ref:Hajkarim2015,noi}.

In \cite{ref:Witten,ref:JH1985}, it has been shown that the quark-hadron PT could result in the formation of some primordial QGP bodies which can,
eventually, survive  up to now. However, in the absence of the baryon chemical potential, the quark content of such bodies (as quark stars)
in the above QCD critical temperature, have limitations and cannot be so large. This is why, incorporating
the quark chemical potential (QCP) into EoS, and owing to a high degree of supercooling at around the same QCD
critical temperature, the possible formation of bodies with larger quark content can be achieved. In  other
words,  neglecting the chemical potential can only be a good approximation for low-density QCD-PT\footnote{
More precisely, although at end of QCD-PT the ratio of quark density number to photon number density
$\eta=\frac{n_q}{n_\gamma}$ may be tiny, of the order of $\simeq10^{-10}-10^{-9}$ (as required by primordial nucleosynthesis),
at temperatures above critical temperature (QGP phase) it is order of unity \cite{ref:j2000} and
\cite{book:cosmo}. The exact value of the baryon asymmetry $\eta$ is released in two independent way, the measurements of the light element abundances based on the Big Bang Nucleosynthesis
(BBN)  and the measurement of the CMB temperature anisotropy. This tiny value shows
that the entropy of the Universe is dominated by a huge margin by the CMB photons since for every baryon in
the Universe there are over 1 billion photons. This means that in QGP phase (the phase above the QCD scale which is expected,
via transition to low phase, to generate stochastic GW(s)) we are dealing with a considerable
number of baryons. Therefore, regarding the possibility
of large quark densities in QGP phase, the approximation of ignoring the QCP cannot be valid and
should be corrected by adding the contribution of the finite QCP. As well as, in
Ref. \cite{plb2013} it is shown that by taking into account the anisotropy of positively and negatively
charged quarks in the early QGP phase, there may have been fluctuations in the chemical potential.}. We note
that the properties of high-density QCD ground state (non-zero baryon chemical potential) play a decisive
role in understanding physics in the core of the neutron stars and heavy-ion collisions \cite{ref:npb2000}.
It is worth noticing that the hypothesis of quark stars have been discussed for the first time by Itoh in \cite{itoh}.
Problems related to dense quark matter and related emissions are discussed in \cite{anderson, flowers, hayashi}.

However, one of great issues of lattice QCD simulations is the inability to predict such models since
the chemical potential effect, when taken into account in the calculations, lead to a complex fermion determinant which
is non-physical. Consequently, in the framework of high-density QCD PT, more attention is  paid to
phenomenological models such as the MIT bag model \cite{ref:mit1974},
and its generalized versions. Given the irrefutable phenomenological role of EoS, it is
interesting to revisit the stochastic GW background power spectrum originated from high-density QCD
first order cosmological PT. The approach consists in improving reliable EoSs with a finite temperature and a chemical potential.
In other words, the chemical potential contributions (that address the possibility of large quark density
bodies in the early Universe) provide a suitable test-bed to investigate on the QCD-based stochastic GW background
from systems with large densities.

This paper is structured as follows. In Sec. \ref{Sec:Eff}, we briefly present a high-density
QCD model, the so called chiral quark model with three flavors (up, down and strange) in which, considering QCP, we deal with the improved EoS.
In Secs. \ref{Sec:SGW1} and \ref{Sec:SGW2}, we show the positive role of the QCP to the detection of stochastic GW background. In Sec.
\ref{Sec:SGW3}, by employing another high-density effective model of QCD, the so called cold QGP
with two light quarks, the up and down quarks, we show that the role of QCP to detect the stochastic signal expected in GWs is highly model-dependent.
Finally,  Sec. \ref{Sec:Disc} is devoted to summary and results. Throughout this work we use natural units
(we set the physical constants $c,~\hbar,~k_B$ and $8\pi G/3$ equal to unity).

\section{A QCD effective model  with finite chemical potential}\label{Sec:Eff}

We perform our investigation in the framework of an effective model which produces improved EoS for QCD dynamics in QGP phase.
By incorporating the QCP, this model provides a developed phenomenological framework to describe high density QCD first order PT in the context
of both high temperature (QGP phase) and quark number density. The main features of the model are briefly review in the following.

\subsection{The chiral quark model}

As first step, let us  present the EoS of the matter in the QGP and hadron phases
\be\label{eos1}
p_{QGP}=\frac{g_{QGP}\pi^2}{90}T^4-V(T)\,, \quad
\rho_{QGP}=\frac{g_{QGP}\pi^2}{30}T^4+V(T)\,,
\ee
and
\be\label{eos1h}
p_{H}=\frac{\rho_{H}}{3}=\frac{g_{H}\pi^2}{90}T^4\,,
\ee
with the QGP and hadronic numbers of degrees of freedoms
$g_{QGP}=37$ and $g_{H}=17.25$, respectively \cite{ref:harko2008, ref:heyd2010, ref:mohsen2013}.
The form of the self-interaction potential $V(T)$
corresponds to a phenomenological model, with the effective Lagrangian given by
\be\label{lag1}
 {\cal L} = \sum_{k=1}^{n_{ f}} \left[ i\bar{\psi}_k \gamma^\mu
\partial_\mu \psi_k - g \bar{\psi}_k (\sigma + i \tau \cdot \pi
\gamma_5) \psi_k \right] +  \frac{1}{2} \partial_\mu \sigma \partial^\mu \sigma
+\frac{1}{2}\partial_\mu \pi \partial^\mu \pi - V(\sigma^2 + \pi^2)\,,
\ee
where the quark fields $\psi_k$ interact with a chiral field, the latter being formed by a $\pi$ meson field plus
the scalar field $\sigma$. The above Lagrangian density can be re-expressed in the  following equivalent form \cite{ref:Kalafatis1992}
\be\label{lag2}
{\cal L} = \sum_{k=1}^{n_{\rm f}} \left[ i\bar{\psi}_k \gamma^\mu
\partial_\mu \psi_k - g \xi (\bar{\psi}_k^{\rm L} U \psi_k^{\rm R} +
\bar{\psi}_k^{\rm R} U^+ \psi_k^{\rm L}) \right]  +\frac{1}{2} \partial_\mu
\xi \partial^\mu \xi + \frac{1}{4} \xi^2\ Tr (\partial_\mu U \partial^\mu U^+) - V(\xi)\,,
\ee
where $U$ denote a component of $SU(2)$ defined by $\xi U = \sigma + i
{\mathbf \tau} \cdot \mathbf{\pi}$ with $\xi = (\sigma^2 + \pi^2)^\frac{1}{2}$, while $\psi_k^{\rm L,R}$
are the left and right-handed elements of the quark field $\psi_k$,
respectively. A generalized self-interaction potential $V(\xi)$ is of the form
\be\label{eff-pot}
 V(\xi)=\frac{1}{2} f_\pi^2 \left(\lambda^2  - \frac{12B}{f_\pi^4}\right)
\xi^2 \left(1  - \frac{\xi}{f_\pi}\right)^2
+B \left[1 + 3\left(\frac{\xi}{f_\pi}\right)^4 - 4\left(\frac{\xi}{f_\pi}\right)^3 \right]\,,
\ee
where the parameters $f_\pi$, $\lambda$, $B$, are related to physical quantities.

The absolute minimum of the self-interaction potential  appears at
$\xi = f_{\pi}$ which, after fitting with the observed pion decay rate, assumes the  numerical value
 $ \xi =93 $ MeV. There is also a local minimum at $\xi=0$ ($V(0)=B$)
corresponding to a false vacuum with the relevant energy density $B$ which is similar to
the perturbative vacuum of the MIT bag model \cite{ref:mit1974}. As a result, the parameter
$B$ in (\ref{eff-pot}) plays the role of bag constant in MIT bag model with numerical values
$B^{1/4}\in(100-200)$ MeV \cite{ref:B}. Now, by adding the  fermion+antifermion
contribution to the above self-interaction potential, we can take into account the finite temperature and
QCP in the model, that is
\be\label{fermion}
\omega_f(T, \mu) = -T\left[\int\!\frac{d^3k}{(2\pi)^3} \ln\left(1 + e^{-(E(k)-\mu)/T}\right) +
\int\!\frac{d^3k}{(2\pi)^3} \ln\left(1 + e^{-(E(k)+\mu)/T}\right)\right]\,,
\ee
with $E(k)=\sqrt{k^2+g^2\xi^2}$. It is worth noticing that in this model the three
quark flavors up, down and strange are fixed, and that the first two are almost massless.
So $g=18$ since $3 \mbox{(quark)}\times 3\mbox{(colore)} \times2\mbox{(spin)}$.
Given that, in the vicinity  $\xi=0$, the role of quark-antiquark pairs assumes a not trivial role.
By expanding the rhs of  Eq. (\ref{fermion}), and maintaining the
strange quark mass $m_s\in(60-170)$ MeV \cite{ref:m}, the self-interaction potential term reads \cite{ref:j2000}
\begin{equation}\label{eff-pot-Tmu}
 V_{T, \mu}(\xi) = B - \alpha_T T^4 -\frac{3\mu^2}{2} T^2 - \alpha_\mu \mu^4 + \gamma_T T^2 + \gamma_\mu \mu^2\,,
 \end{equation}
 \[
 \alpha_T = \frac{7\pi^2}{20}, \quad \alpha_\mu = \frac{3}{4\pi^2}, \quad
\gamma_T = \frac{m_s^2}{4} , \quad \gamma_\mu = \frac{3m_s^2}{4\pi^2}\,.
\]
The point that should be noted is that in (\ref{eff-pot-Tmu}), $\mu$, unlike $B$ and $m_s$,  is not simply
a parameter but it is connected to the quark number density $n_q=3\mu T^4+4\alpha_{\mu} \mu^3
-2\gamma_\mu \mu^2$. In its absence, the high-density QCD converts to low-density QCD.
As a result, the final form of EoS (\ref{eos1}) can be re-expressed as follows\footnote{Here, the additional terms appearing in the self-interaction potential $V(T)$ (more exactly, $V(\xi)$) come from the non-linear interactions of the gluons (due to the chiral field $\xi$). In other words, by considering $V(T)$ rather than $B$ in standard MIT bag model (including free gluons), one arrives at the more general EoS (\ref{eos2a}) and (\ref{eos2b}) which, unlike its standard counterpart, are not related to each other through the usual Legendre transformation $\rho=T\frac{dP}{dT}-p$.}

\bea\label{eos2a}
p_{QGP} &=& \left(\frac{37\pi^2}{90}+ \alpha_T\right)T^4 +\left(\frac{3\mu^2}{2}-\gamma_T \right)T^2 + \alpha_\mu \mu^4 - \gamma_\mu \mu^2-B\,, \\
   & & \nonumber \\
\rho_{QGP}&=& \left(\frac{37\pi^2}{30}- \alpha_T\right)T^4 -\left(\frac{3\mu^2}{2}-\gamma_T \right)T^2 - \alpha_\mu \mu^4 + \gamma_\mu \mu^2 +B\,, \label{eos2b}
\eea
with the chemical potential $\mu \equiv \mu_{u}=\mu_{d}=\mu_{s}$.
Clearly  ignoring the temperature and chemical potential
effects in self-interaction potential, we obtain the well-know EoS of MIT
bag model \cite{ref:mit1974}. Using the condition $p_{QGP}(T_c)=p_{H}(T_c)$
\cite{ref:prd1986} in this improved model, the critical temperature $T_{c}$ turns out to be
\be\label{Tc1}
T_{c}=\left[\frac{\frac{3\mu^2}{2}-\gamma_T }{\frac{\pi^2}{45}\Delta g+ 2\alpha_T}\, \left(-1\pm \sqrt{1-\frac{\big(\frac{2\pi^2}{45}\Delta g+ 4\alpha_T\big)\big(\alpha_\mu \mu^4 - \gamma_\mu \mu^2-B \big)}{(\frac{3\mu^2}{2}-\gamma_T )^2}}\right)\right]^{1/2}\,,
\ee
where $\Delta g=19.75$. In the above relation, the negative sign solution leads to a
physical critical temperature provided that $\mu<\frac{m_s}{\sqrt{6}}$.

\section{QCD-Based SGW spectrum revisited by chemical potential}\label{Sec:SGW1}

\begin{figure}
\begin{center}
\begin{tabular}{c}\hspace{-0.5cm}\epsfig{figure=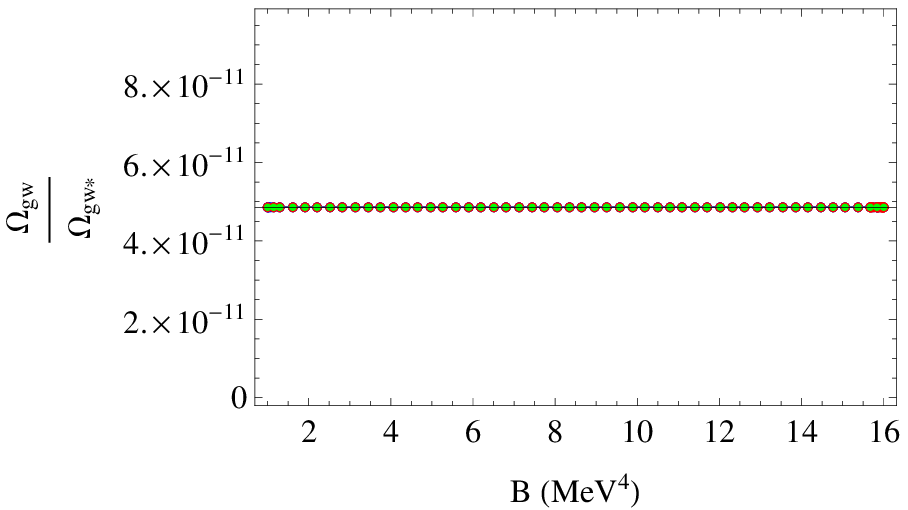, width=3in,height=2in,angle=0}
\hspace{0.5cm} \epsfig{figure=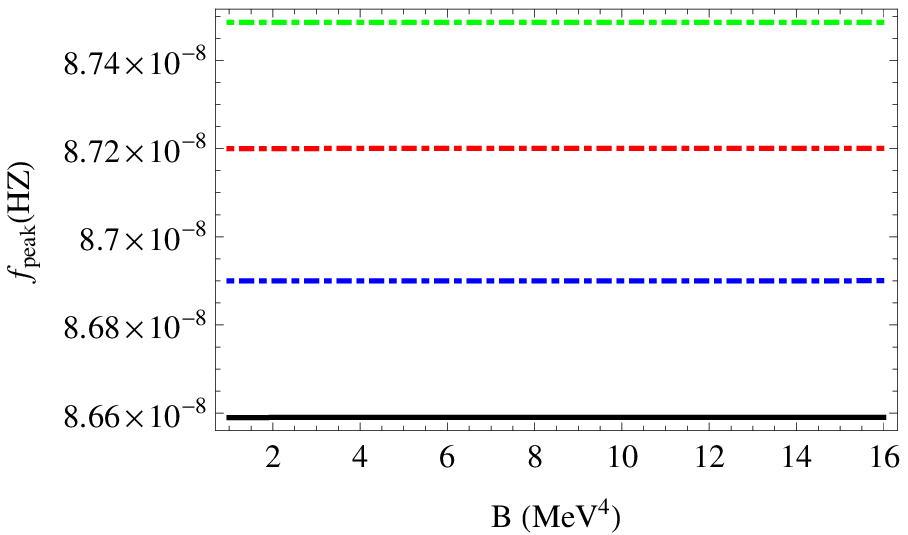, width=3in,height=2in,angle=0}
\end{tabular}
\end{center}
\caption{The fractional energy density of GW signal and the received frequency at the current time
in terms of bag constant for the chiral quark model with $\mu$ variable: $\mu=0$ (black solid) $\mu=10$
(blue dot-dashed), $\mu=20 $ (red dot-dashed), $\mu=30$ (green dot-dashed), from left to right panels.
The axes are re-scaled by the factor $10^{-8}$.}
\label{BC}
\end{figure}

Generally, the propagation of the GWs from the of QCD-PT until today allows to estimate, in principle, the present observable GW background.
With the assumption that since the PT the Universe expanded adiabatically, the entropy per comoving volume $S \propto V\, g_s\, T^3$ remains constant
($\dot S/S=0\,$), so that the temperature variation with respect to time  has the following form
\be \label{Tdot}
\frac{dT}{dt} = -H T\,\left(1+ \frac{T}{3g_s}\frac{dg_s}{dT} \right)^{-1}\,,
\ee
where $V=a^3$ with $a$ the scale factor and $g_s$ coming from the effective number of freedom degrees involved in the entropy density.
The Hubble parameter $H$ in the above expression allows to rewrite Eq. (\ref{Tdot}) in terms of scale factor and energy density of the GWs
\be\label{scale}
\frac{a_*}{a_0} = \exp \left[\int_{T_*}^{T_0} \frac{1}{T} \left(1+ \frac{T}{3g_s}\frac{dg_s}{dT}\right)\, dT\right]\,,
\ee
and
\be\label{eq:rho-gw}
\rho_{\rm gw}(T_0) = \rho_{\rm gw}(T_*)\,\exp\left[\int_{T_*}^{T_0} \frac{4}{T}\left(1+ \frac{T}{3g_s}\frac{dg_s}{dT}\right)\, dT\right]\,,
\ee
respectively. Here $\rho_{\rm gw}(T_0)$ and $\rho_{\rm gw}(T_*)$ represent
the energy density of GWs at current and PT epochs, respectively. It is worth noticing that Eq. (\ref{eq:rho-gw}) comes from
the Boltzmann equation, $\frac{d}{dt}(\rho_{gw}\, a^4) = 0$  as a result of the fact that GWs decouple from rest of the Universe. The today GW density
parameter is defined as (up to the end of this paper, the indexes  `` 0 '' and `` * '' address the quantities at today and PT eras, respectively)
\be\label{eq:gw_den}
\Omega_{\rm gw}=\Omega_{\rm gw*}\bigg(\frac{H_*}{H_0}\bigg)^2 \exp\bigg[\int_{T_*}^{T_0}
  \frac{4}{T}\left(1+\frac{T}{3g_s}\frac{dg_s}{dT}\right)\, dT\bigg]\,,
\ee
where
 \[
 \Omega_{\rm gw} = \frac{\rho_{\rm gw}(T_0)}{\rho_{\rm cr}(T_0)}\,, \quad\quad
 \frac{\rho_{\rm cr}(T_*)}{\rho_{\rm cr}(T_0)} =\left(\frac{H_*}{H_0}\right)^2\,.
 \]
In what follows, using the continuity equation, $\dot\rho = -3H\rho \big(1+ w_{eff}\big)$
(here $\rho (p)$ being the total energy (pressure) density of the Universe
and $w_{\rm eff}=\frac{p}{\rho}$ denotes the effective EoS parameter),
we can derive the ratio of the Hubble parameter during PT up
to its today value. To this end, we have to determine the energy density
at  PT epoch. It is obtained by mixing eq. (\ref{Tdot}) into the
continuity equation and integrating from some early time in the
radiation dominated era with  temperature $T_r$ till the PT epoch, i.e.
\be \label{eq:energy}
\rho(T_*) = \rho(T_r)\, \exp\bigg[\int_{T_r}^{T_{*}}  \frac{3}{T}\,(1+ w_{\rm eff} )
  \left(1+ \frac{T}{3g_s}\frac{dg_s}{dT}\right)\, dT\bigg]\,.
\ee
Now, with the above result at hand, as well as the relation $H_*^2 = \rho_*$, we have
\be \label{eq:H*}
\bigg(\frac{H_*}{H_0}\bigg)^2 = \Omega_{r0} \bigg(\frac{a_0}{a_r}\bigg)^4  \exp\bigg[\int_{T_{r}}^{T_{*}}
\frac{3(1 + w_{\rm eff})}{T}\left(1+ \frac{T}{3g_s}\frac{dg_s}{dT}\right)dT\bigg]\,,
\ee
where $\Omega_{r0}$ denotes the today value of fractional energy density
of radiation with numerical value $\Omega_{r0}\simeq 8.5\times10^{-5}$
and $T_r = 10^4$ GeV (this value of the temperature is above QCD era, in particular the EW era. Given that the EW force freeze out at $T\simeq 10^{15}$K $\simeq  10^2$GeV, the value of $T_r$ should be fixed at $T_r>10^2$GeV. Notice that even fixing a larger value of the temperature $T_r$, our results do not change). As a consequence, the GW spectrum measured today, Eq. (\ref{eq:gw_den}), reads
\be \label{eq:GW}
\Omega_{\rm gw} = \Omega_{r0} \Omega_{\rm gw*}  \exp\bigg[\int_{T_*}^{T_r}
  \frac{4}{T'}\left(1+ \frac{T}{3g_s}\frac{dg_s}{dT}\right)\, dT\bigg] \,
  \exp\bigg[\int_{T_{r}}^{T_{*}} \frac{3}{T}\,(1+w_{\rm eff})
  \left(1+ \frac{T}{3g_s}\frac{dg_s}{dT}\right)\, dT\bigg]\,.
\ee
The above relation can be rewritten in the straightforward form
\be \label{eq:smple}
\frac{\Omega_{\rm gw}}{\Omega_{\rm gw*}}=\left(\frac{g_{s}(T_0)}{g_{s}(T_*)}
\right)^{4/3} \,\frac{T_{0}^{4}}{T_{*}^{4}}\, \frac{H_{*}^{2}}{H_{0}^{2}}\,,
\ee
where $H_0^2=\rho_{cr}\simeq 8\times10^{-35}~MeV^4$.
The GW frequency peak, redshifted to the current epoch\footnote{We have to consider the ratio between the
frequency received by an observer at present time to that emitted during the transition.}, is
\be \label{eq:frequency}
f_{peak}=f_{*}\left(\frac{a_*}{a_0}\right)= \left(\frac{g_{s}(T_0)}{g_{s}
(T_*)}\right)^{1/3}T_{0}\,\frac{\rho^{1/2}(T_*)}{T_{*}}\,,
\ee
where $f_*=H_*=\rho_*^{1/2}$. Now for displaying the phenomenological feedback of (\ref{eq:smple}) and (\ref{eq:frequency}), we have
\bea \label{eq:sample1}
\frac{\Omega_{\rm gw}}{\Omega_{\rm gw*}} &=& \bigg(\frac{g_{s}(T_0)}{g_{s}
(T_{c})}\bigg)^{4/3}\frac{g \pi^2T_{0}^4}{24\times 10^{-34}\text{MeV}^4}\,
\bigg[1- \alpha_T+\frac{2\gamma_T -3\mu^2}{2T_{c}^2}
+\frac{B+\gamma_\mu \mu^2- \alpha_\mu\mu^4}{T_{c}^4}\bigg]\,, \\
 & & \nonumber \\
f_{peak} &=& T_{0}\bigg(\frac{g_{s}(T_0)}{g_{s} (T_{c})}\bigg)^{1/3}\, \frac{\rho_{QGP}^{1/2}(T_{c})}{T_{c}}\,. \label{eq:fre1}
\eea
By setting the numerical value\footnote{Note that the general behavior of
plots in Fig. \ref{BC} is independent of allowed values for $m_s$.}
$m_{s}=150$ MeV, we can see the effect of $\mu$-terms
on the fractional energy density of GWs as well as the frequency peak received
at the current time in Fig. \ref{BC}.

As we can see from the black to green lines,
the frequency peak, red-shifted to the present time, grows by increasing the values of the chemical potential, although the fractional energy density of GW signal remains unchanged. In the next Section we shall discuss the QCP effect on the stochastic GW background at present epoch.

\section{Two QCD sources of GWs: bubble wall collisions and magnetohydrodynamic
 turbulence} \label{Sec:SGW2}

Concerning the stochastic  GW generators acting in QCD first order PT, we can consider
 \emph{"bubble wall collisions"} (BWC) \cite{ref:Kosowsky:1991ua, ref:Kosowsky:1992rz, ref:Caprini:2007xq, ref:Huber:2008hg}
and \emph{" turbulent magnetohydrodynamic (MHD) "} \cite{ref:Caprini:2006jb, ref:Caprini:2015zlo}.
These mechanisms  can contribute to clarify the role of QCP in effective QCD models for the
detection of stochastic GW spectrums. More exactly, we can obtain a rough estimate
of the GW amplitude $\Omega_{gw}h^{2}$ in the presence of QCP to illustrate its
phenomenological role. To do this end, we focus our attention on $\Omega_{gw*}\equiv\Omega_{gw}(T_{*})$
which  is necessary to determine  the amplitude of stochastic GW background signal. The above mentioned two sources are
components of the bubble percolation which occurs after bubble nucleation and bubble expansion in the QCD
first order PT. Historically, Witten\footnote{This idea, later by Hogan, has been extended to the case of
electro-weak PT \cite{ref:Hogan}.} proposed for the first time  the idea of detectable
QCD-generated GWs from violent BWC as well as the turbulent motion of bulk fluid
as remnant of BWC \cite{ref:Witten}.  In \cite{ref:prd1994}, such a process was first estimated as a Kolmogorov spectrum
under quadrupole approximation.  For a detailed review see \cite{ref:Caprini:2015zlo, ref:rev1}.

\begin{figure}
\begin{center}
\begin{tabular}{c}\hspace{-0.5cm}\epsfig{figure=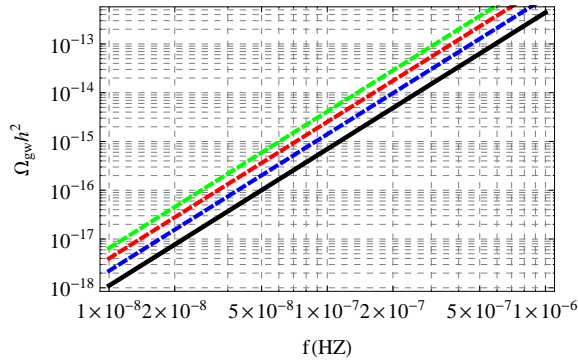, width=3in,height=2in,angle=0}
\end{tabular}
\end{center}
\caption{Double-logarithmic plot of the amplitude of QCD-based GW signal spectrum arising from the contribution of BWC plus MHD
turbulence in framework of chiral quark model versus frequency. By fixing $B^{1/4}=100$ MeV and $m_s=150$ MeV and $n=10$, we
use $\mu$ variable: $\mu=0$ (black solid), $\mu=20$ MeV (blue dashed), $\mu=40$ MeV
(red dashed), $\mu=60$ MeV (green dashed). }
\label{sm1}
\end{figure}

Numerical simulations, employing the envelope approximation \footnote{As recently reported in \cite{Cutting2018} by
using the large scale numerical lattice simulations, it has been shown that the rate of fall off of the
electro-weak PT-based GW spectrum in high frequencies, is slightly faster than what is expected 
by numerical simulations using the thin-wall envelope approximation
(i.e. as $f^{-1.5}$ rather than $f^{-1}$).  This means that the peak of the power spectrum
arising from BWC process, is slightly moving towards lower frequencies from that of the envelope
approximation. It is guessed that the origin of this deviation is due to the lack of calculation
of some overlap regions of the bubbles in envelope approximation. It is interesting to mention
that before releasing results in \cite{Cutting2018}, envelope approximation was in a good agreement with 
lattice simulations \cite{Weir} and also with recent analytical models \cite{Jinno}.} \cite{ref:Huber:2008hg} for
BWC process related to the Kolmogorov-type turbulence for MHD turbulence  process \cite{ref:K-type,ref:Bohe2012, ref:Cap2009},
suggest that  contributions to the GW spectrum can be read as
\bea \label{bub}
h^2\Omega_{\rm gw*}^{(bwc)}(f) &=& \bigg(\frac{H_*}{\beta}\bigg)^2
\bigg(\frac{ \kappa_b\delta}{1+\delta}\bigg)^2
\bigg(\frac{0.11 u^3}{0.42 + u^2}\bigg)\,\frac{3.8\,(f/f_{bwc})^{2.8}}{1 + 2.8\,(f/f_{bwc})^{3.8}}\,, \\
 & & \nonumber \\
h^2\Omega_{\rm gw*}^{({\rm mhd})}(f) &=& \,\bigg(\frac{H_*}{\beta}\bigg)\,
\bigg(\frac{\kappa_{{\rm mhd}}\,\delta}{1+\delta}\bigg)^{3/2}\,
\frac{u(f/f_{{\rm mhd}})^3}{(1 +f/f_{{\rm mhd}})^{11/3}\, (1 + 8\pi\,f/{\cal H}_*)}\,, \label{mhd}
\eea
with
 \[
f_{bwc} = \frac{0.62 \beta}{(1.8-0.1 u + u^2)}\,\left(\frac{a_*}{a_0}\right)\,, \quad
f_{{\rm mhd}} = \frac{7\beta}{4u} \left(\frac{a_*}{a_0}\right)\,, \quad
{\cal H}_* = H_*\left(\frac{a_*}{a_0}\right)\,,
 \]
where $f_{bwc}$ and $f_{{\rm mhd}}$ are the today peak frequency of the stochastic GWs produced by BWC and MHD contributions
during PT, respectively. Here the parameters $\kappa_b$, $u$, $\beta$,
$\delta$ and $\kappa_{{\rm mhdt}}$ represent
the fraction of the latent energy related to first order PT residues on the bubble wall,
the wall velocity, the characteristic time-scale of the PT, the ratio of the vacuum energy density
released  during PT relative to that of the radiation and the fraction of latent heat
energy generated in turbulence regime, respectively. Using eq. (\ref{scale})
and setting $H_{*}=\rho_*^{1/2}(T_{*})$, $T_*=T_{c}$ and $\beta=nH_*$,
the above equations can be re-expressed as
\bea \label{bub-final}
h^2\Omega_{\rm gw*}^{(bwc)}(f) &=& \bigg(\frac{n \kappa_b\delta}{1+\delta}\bigg)^2
\bigg(\frac{0.11 u^3}{0.42 + u^2}\bigg) \chi_{bwc}\,, \\
h^2\Omega_{\rm gw*}^{({\rm mhd})}(f) &=& \,\bigg(\frac{\kappa_{{\rm mhd}}\,\delta}{1+\delta}\bigg)^{3/2}\, \chi_{\rm mhd}\,, \label{mhd-final}
\eea
with
\be
\chi_{bwc}= \frac{3.8\left(\frac{(100u^2-10u+180)f}{62n}~\left(\frac{g_{s}(T_0)}{g_{s}
(T_{c})}\right)^{-1/3}\, \frac{T_{c}}{T_{0}\sqrt{\rho_{QGP}(T_{c})}}\right)^{2.8} }{1+2.8\left(\frac{(100u^2-10u+180)f}{62n}~
 \bigg(\frac{g_{s}(T_0)}{g_{s} (T_{c})}\bigg)^{-1/3}\, \frac{T_{c}}{T_{0}\sqrt{\rho_{QGP}(T_{c})}}\right)^{3.8}}\,,
\ee
and
\bea
\chi_{\rm mhd}&=& \chi_{0} \, \left[\bigg(1+\frac{fu}{1.75n} \bigg(\frac{g_{s}(T_0)}{g_{s}(T_{c})}\bigg)^{-1/3}\frac{T_{c}}
{T_{0}\sqrt{\rho_{QGP}(T_{c})}}\bigg)^{11/3}
\bigg(1+8\pi f \bigg(\frac{g_{s}(T_0)}{g_{s}(T_{c})}\bigg)^{-1/3}\frac{T_{c}}
{T_{0}\sqrt{\rho_{QGP}(T_{c})}}\bigg)\right]^{-1}\,, \\
\chi_{0}&\equiv &\frac{f^3u^4}{5.36n^2}~ \frac{g_{s}(T_{c})}{g_{s}(T_0)}~\frac{T_{c}^{3}} {(T_{0}\sqrt{\rho_{QGP}(T_{c})})^3}\,, \nonumber
\eea
respectively. By inserting $h^2\Omega_{\rm gw*}=h^2\big(\Omega_{\rm gw*}^{(bwc)}(f)+\Omega_{\rm gw*}^{({\rm mhd})}(f)\big)$ into
Eq. (\ref{eq:smple}) and using Eq. (\ref{Tc1}), we  obtain a rough estimate of the GW amplitude
$\Omega_{gw}h^{2}$ for two effective QCD models at hand, see Fig. \ref{sm1}.
As can be seen in Fig. \ref{sm1}, by increasing the values of $\mu$ (from black to green),
the height of the stochastic GW signal becomes larger. For the today peak frequency arising
from BWC and MHD contributions, we  have
\be\label{f-total}
f_{total}=n T_{0}\bigg(\frac{62}{100u^2-10u+180}+\frac{7}{4u}\bigg) \, \bigg(\frac{g_{s}(T_0)}{g_{s}(T_{c})}\bigg)^{1/3}
\frac{\rho_{QGP}^{1/2}(T_{c})}{T_{c}}\,,
\ee
where, by fixing the relevant values for $n$ and $u$, one can  show that, in agreement with Fig. \ref{BC}, the
peak frequency, redshifted to present time, grows by increasing $\mu$. From the perspective of phenomenology, the two quantities $f_{peak}$ and $\Omega_{gw}h^{2}$
provide the chance of detecting the stochastic GWs in future observations.

\begin{figure}
\begin{center}
\begin{tabular}{c}\hspace{-0.5cm}\epsfig{figure=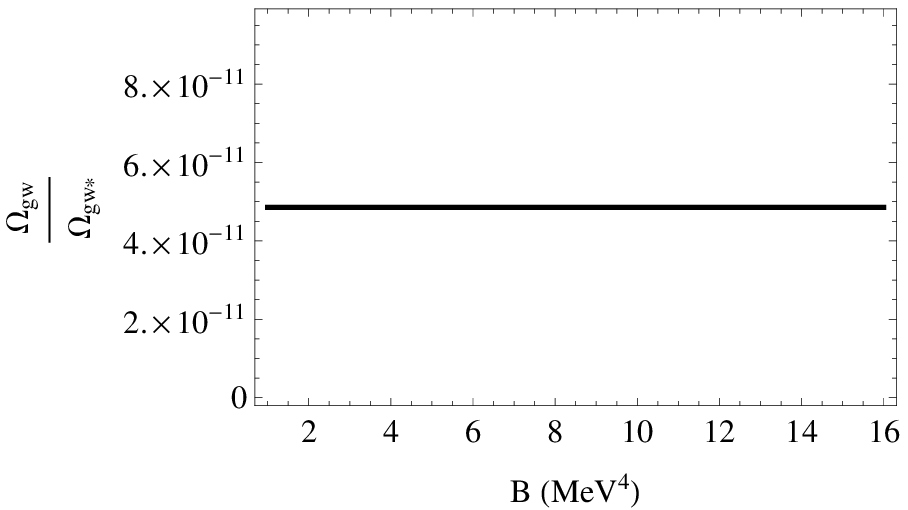, width=3in,height=2in,angle=0}
\hspace{0.5cm}\epsfig{figure=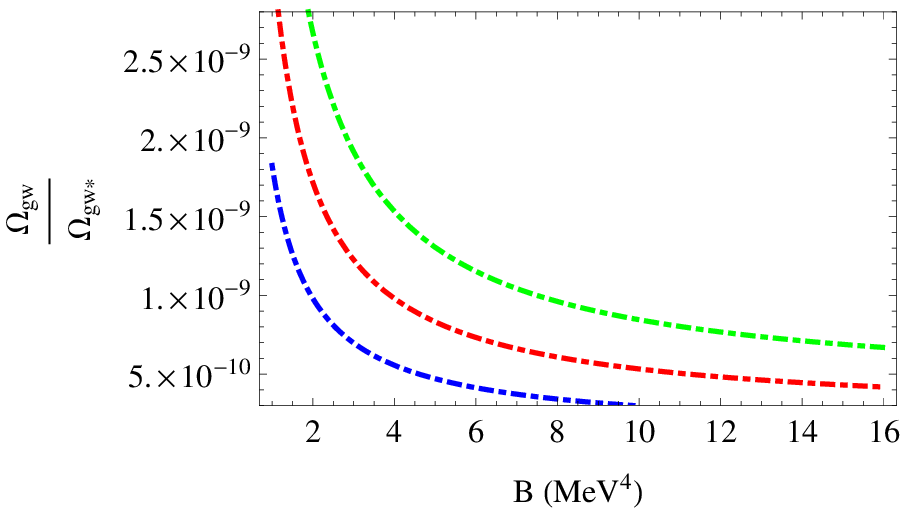, width=3in,height=2in,angle=0}\\
\hspace{0.5cm}\epsfig{figure=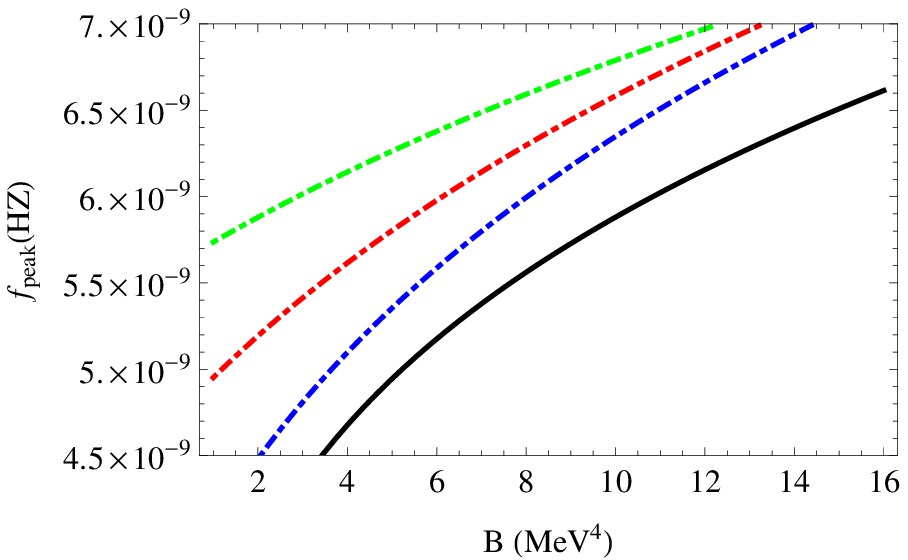, width=3in,height=2in,angle=0}
\end{tabular}
\end{center}
\caption{The fractional energy density of GW signal (upper two  panels) and the peak
frequency red-shifted to the current time (lower panel) in terms of bag constant for
cold QGP model with $\mu$ variable: $\mu=0$ (black solid), $\mu=10$ MeV (blue dot-dashed),
$\mu=15$ MeV (red dot-dashed), $\mu=20$ MeV (green dot-dashed). The axes are re-scaled by a
factor $10^{-8}$.}
\label{signal}
\end{figure}

\section{The model-dependent QCP contribution to detect the
stochastic GW background}\label{Sec:SGW3}

The role of QCP for the  detection of stochastic GWs background is model-dependent.
This statement can be demonstrated by considering a phenomenological
high-density QCD model with two flavor quarks (up, down) known as \emph{"cold QGP"}. However,  it is possible to show that also  the strange quark can contribute to the above
phenomenology.

This QCD model  has been considered for the first time  in \cite{ref:plb2011} and then it was extended within
neutron stars for investigating the cold and high dense quark gluon plasma
in the inner core \cite{ref:prd2012}. The starting point in
\cite{ref:plb2011} is focused on the QCD Lagrangian density where,
using a relativistic mean field approximation, the gluon field splits
into two modes with low (soft) and high (hard) momenta.
The low momentum modes are re-expressed in terms of the gluon
condensate which results in a residual non-vanishing value
in the QGP phase. By repeating the steps of  finite temperature
formalism proposed in \cite{ref:prc1990} for the effective Lagrangian
density of \cite{ref:plb2011}, the following expressions
for the pressure and the energy density can be derived \cite{ref:npa2015}
\begin{eqnarray}\label{p}
p_{cold-QGP} &=& {\frac{3\pi\alpha_s}{4{m_{g}}^{2}}}{n_q}^{2}-B+ \sum_{f}\,{\frac{\gamma_{f}}{6\pi^{2}}}\int_{0}^{\infty} dk \,\frac{{k}^{4}}
{\mathcal{E}_{f}}\, \Big(d_{f}+\bar{d}_{f} \Big)
+ {\frac{\gamma_g}{6\pi^{2}}}\int_{0}^{\infty} d{k}\frac{k^{2}}{(e^{k/T}-1)}\,, \\
 & & \nonumber \\
\rho_{cold-QGP} &=& {\frac{3\pi\alpha_s}{4{m_{g}}^{2}}}{n_q}^{2} + B + \sum_{f}\,{\frac{\gamma_{f}}{2\pi^{2}}} \int_{0}^{\infty} dk \, k^{2}\, {\mathcal{E}_{f}}\,\Big(d_{f}+ \bar{d}_{f}\Big) + {\frac{\gamma_g}{2\pi^{2}}}\int_{0}^{\infty} d{k}\frac{k^{2}}{(e^{k/T}-1)}\,, \label{e}
\end{eqnarray}
where
 \[
\mathcal{E}_{f}=\sqrt{m_{f}^{2}+k^{2}}\,, \quad
d_{f}\equiv \frac{1}{1+e^{(\mathcal{E}_{f}-\nu_{f})/T}}\,, \quad
\bar{d}_{f}\equiv \frac{1}{1+e^{(\mathcal{E}_{f}+\nu_{f})/T}}\,,
\]
denote the energy of the quark of flavor $f$ and the Fermi distribution
functions\footnote{Concerning the Fermi gas in Fermi distribution function, the Fermi
energy is proportional to the density which is due to the chemical potential dependence on it. In other words,
the chemical potential  is proportional to the  density.} with the chemical potential $\nu_{f}$.
Taking two light quark flavors u and d, with the same masses, and fixing the high temperature regime given
by $T\gg \nu_{f}$, $T\gg m_{f}$ and $\mathcal{E}_{f}/T>\nu_{f}/T$, Eqs. (\ref{p}) and (\ref{e}) become \cite{ref:npa2015}
\bea\label{pf}
p_{cold-QGP} &=& \bigg({\frac{37\pi^{2}}{90}}+{\frac{3\pi\alpha_s\mu^{2}}{4{m_{g}}^{2}}}\bigg){T^{4}} +{\frac{{\mu}^{2}}{2}} \,T^{2}\,-B\,, \\
 & & \nonumber \\
\rho_{cold-QGP}&=& \bigg({\frac{37\pi^{2}}{30}}+{\frac{3\pi\alpha_s\mu^{2}}{4{m_{g}}^{2}}}\bigg){T^{4}} +{\frac{3{\mu}^{2}}{2}} \,T^{2}\,+ B\,, \label{ef}
\eea
with the chemical potential $\mu \equiv \nu_{u}=\nu_{d}$, while $\alpha_s<1$ is a strong coupling constant arising from the coupling between
the quarks and hard gluons (the number of degrees of freedom $37$ appearing in above EoS corresponds to $\alpha_{s}=0.5$ \cite{ref:BookQCD}).
Note that, the  high temperature condition $T\gg m_{f}$ leads to the derivation of the above analytical solutions.
Furthermore, the above solutions are obtained assuming the statistical factors
$\gamma_f=2\textrm{(spins)}\times 3\textrm{(colors)}=6$ and $\gamma_g=2\textrm{(polarizations)}\times 8\textrm{(colors)}=16$ for each quark
and gluons, respectively. Eqs. (\ref{pf}) and (\ref{ef}) are reliable at high density QGP phase with finite temperature and chemical potential
so that, in case of fixing  $\mu=0$, they come back to their simple counterparts of the MIT bag model \cite{ref:mit1974}. With the same previous
condition, the critical temperature is
\be\label{Tc2}
T_{c}=\left[\frac{3\mu^{2}}{\frac{2\pi^{2}}{45}\Delta g+\frac{3\pi\alpha_s\mu^{2}}{{m_{g}}^{2}}}
\left(-1\pm\sqrt{1+\frac{16B\left(\frac{\pi^{2}}{90}\Delta g+\frac{3\pi\alpha_s\mu^{2}}{{4m_{g}}^{2}}
\right)}{9\mu^4}}\right)\right]^{1/2}\,,
\ee
where in the limit $\mu\rightarrow0$, Eq. (\ref{Tc2}) reduces to the usual expression of MIT bag model, i.e. $T_{c}=\displaystyle{\left(\frac{90B}{\pi^2\Delta g}\right)^{1/4}}$. Here, without constraints on the  $\mu$ value, the solution with positive sign in Eq. (\ref{Tc2}) is acceptable and physically meaningful.
Equations (\ref{eq:smple}) and (\ref{eq:frequency}) become
\bea \label{eq:sample2}
\frac{\Omega_{\rm gw}}{\Omega_{\rm gw*}} &=& \bigg(\frac{g_{s}(T_0)}{g_{s} (T_{c})}\bigg)^{4/3}
\frac{g \pi^2T_{0}^4}{30H_0^2}\,
\bigg[1+\frac{3\alpha_s\mu^2}{16m_g^2}+\frac{3\mu^2}{2T_{c}^2}+\frac{B}{T_{c}^4}\bigg]\,, \\
 & & \nonumber \\
f_{peak} &=& T_{0}\bigg(\frac{g_{s}(T_0)}{g_{s}
(T_{c})}\bigg)^{1/3}\,\frac{\rho_{cold-QGP}^{1/2}(T_{c})}{T_{c}}\,.  \label{eq:fre2}
\eea
\begin{figure}
\begin{tabular}{c}\hspace{-0.5cm}\epsfig{figure=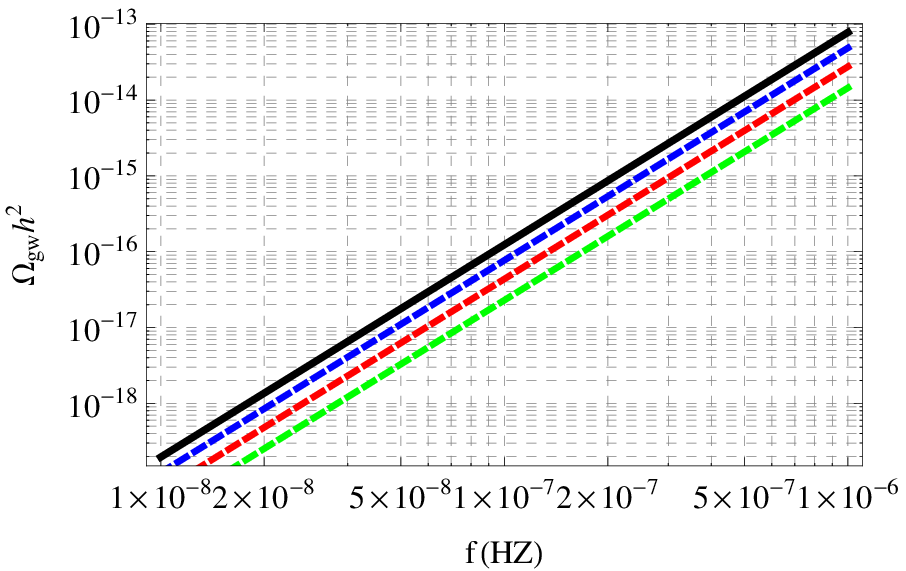, width=3in,height=2in,angle=0}
\epsfig{figure=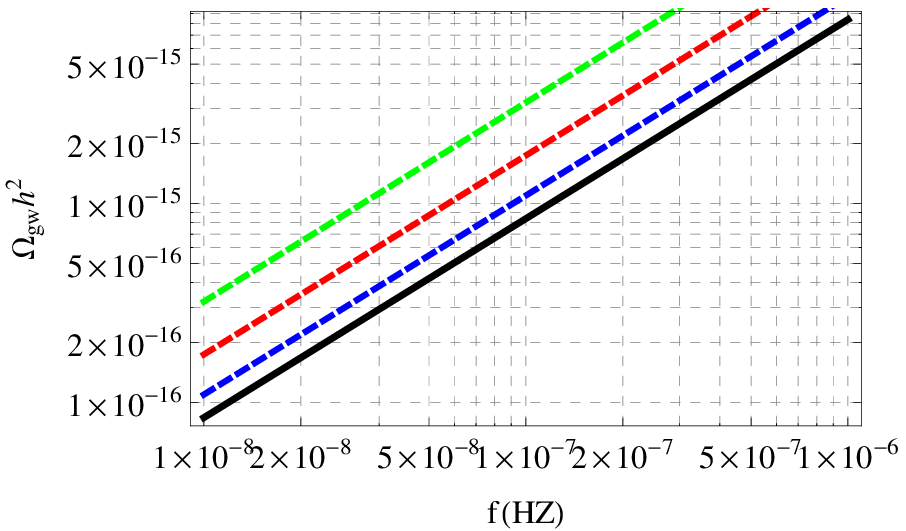, width=3in,height=2in,angle=0}
\end{tabular}
\caption{Double-logarithmic plot of the amplitude of high-density QCD-based GW signal arising from the contribution
of BWC+ MHDT in framework of cold QGP model (left panel) and chiral quark model with
$\gamma_T=0=\gamma_{\mu }$ (right panel) versus frequency. By fixing $B^{1/4}=100$ MeV, $m_g=10$ Mev,
$n=10$ and $u=0.7$, we use $\mu$ variable: $\mu=0$ (black solid), $\mu=20$ MeV (blue dashed),
$\mu=40$ MeV (red dashed), $\mu=60$ MeV (green dashed), respectively.}
\label{sm2}
\end{figure}
The effect of the chemical potential on the fractional energy density of GWs and the peak frequency
received at the present time are  qualitatively displayed in Figs. \ref{signal},
assuming $m_g=10$ MeV for dynamical gluon mass as well as other
numerical values for parameters involved in cold QGP model. Figures show that by increasing the values of
chemical potential (from black to green) the height of the fractional energy
density of GW signal as well as the peak frequency redshifted to the current time
become larger. However, we see in Fig. \ref{sm2},  in contrast to what is represented
in previous high-density QCD effective model, as soon as $\mu$ increases, the amplitude of
stochastic GW background $\Omega_{gw}h^{2}$ falls with respect to the case $\mu=0$.
One can immediately see that the behaviors in
Figs. \ref{sm1} and \ref{sm2} (top panel) change owing to the absence of quark strange
contribution in the cold QGP model. The absence of  contribution of quark strange  is
related to the values of  parameters  $\gamma_T=0=\gamma_{\mu }$.
Finally,  given  the essential role of EoS
in phenomenology aspects of QCD, one can realize that the difference in  predictions
is related to the different EoS(s) related to  the two models.

\section{Summary and discussion }\label{Sec:Disc}
We have explored the implication of high-density QCD first order PT
in the production of  stochastic GW background. We started our considerations
assuming an effective QCD model with three chiral  quark flavors (up, down,
strange) including finite temperature and chemical potential. In particular,
by focusing on two measurable quantities, namely  the peak frequency $f_{peak}$ redshifted
at today time and the fractional energy density $\Omega_{gw}h^2$, we have
revisited the stochastic background of GW spectrum propagated from first
order PT to the QCD era. We have assumed  a high-density regime  with a finite quark
chemical potential. It  turns out an increasing of the characteristic
frequency and the amplitude of stochastic GW signal received today ,
as the chemical potential increases (Figs. \ref{BC} (right panel) and \ref{sm1}).
This can be considered  a solid feedback because, in the presence of  chemical
potential, the chance of measuring the stochastic background of GWs, caused by the QCD-PT, could be a reliable goal in
future observations by  experiments like the  ``Pulsar Timing Array'' (PTA) and the ``Square Kilometer Array'' (SKA).
Due to the reinforcing role of quark chemical potential, this signal amplification enhances the possibility of
locating the stochastic GWs emerged from QCD-PT into the sensitivity range of SKA/PTA.  Even if  this possibility is weak at the moment, by updating the sensitivity of the related experiments in the future,
we can still remain hopeful. 

As a final comment, we discuss the possibility that the constructive contribution of the quark chemical potential
to detection of stochastic GWs background could be model-independent (i.e. the existence
of a chemical potential independent on the model under consideration might imply an amplification of the GWs due to the QCD-PT).
To investigate this topic, we have extended our study to cold QGP effective model of QCD first order PT with two light quarks: up, down.
Here, despite to the previous result for $f_{peak}$, it is possible to show that $\mu$ increases the amplitude of stochastic GW background
$\Omega_{gw}h^{2}$, but falls with respect to the case $\mu=0$ (Figs. \ref{signal} (bottom panel) and \ref{sm2} (left panel)). So, the underlying
effective QCD model predicts stochastic GWs with the peak frequency higher, but with amplitude signal weaker than the one corresponding to $\mu=0$.
An important point that should be noted is that in contrast with Lattice outputs which addresses crossover PT,
here both the effective QCD models, also in absence of the quark chemical potential, result in a first order PT. Therefore, the appearance of the GW signal
in these two models for the case  $\mu=0$ is not unexpected. Of course, taking into account the finite chemical potential into Lattice simulation, there
is also the possibility for first order QCD-PT, see \cite{ref:prd2013} for instance.
As a consequence, these feedbacks suggest that the contribution of the quark (baryon) chemical potential to the detection of
stochastic GW background is highly  model-dependent. The benefit of the model-dependent output of chemical potential is that, by tracking
the effective QCD models in light of GW spectrum, future GW observations can fix the dynamics of the QCD-PT.
More precisely, the stochastic GW spectrum can be regarded as a criterion for classifying the effective QCD models from a phenomenological point
of view.

\section*{Acknowledgments}
SC  and GL acknowledge the support of  INFN ({\it iniziative specifiche}  QGSKY and TEONGRAV).
This paper is based upon work from COST action CA15117 (CANTATA), supported by COST (European
Cooperation in Science and Technology). The work of M.Kh. has been financially supported by
Research Institute for Astronomy and Astrophysics of Maragha (RIAAM) under research project
No. 1/5750-12.

\end{document}